\documentclass[aps,prl,preprint, preprintnumbers,amsmath,amssymb,showpacs,superscriptaddress,amsfonts]{revtex4-1}

\pdfoutput=1

\usepackage{graphicx}
\usepackage{amsmath}
\usepackage{amstext}
\usepackage{amssymb}
\usepackage{amsfonts}
\usepackage{amsbsy}
\usepackage{dcolumn}
\usepackage{bm}
\usepackage{multirow}

\usepackage{hyperref}

\usepackage[bbgreekl]{mathbbol}
\usepackage{amscd}

\newcommand{\beq}{\begin{equation}}
\newcommand{\eeq}{\end{equation}}

\newcommand{\bsi}{{\boldsymbol{I}}}

\newcommand{\ie}{{\emph{i.e.~}}}
\makeatletter

\newcommand{\Rmnum}[1]{\expandafter\@slowromancap\romannumeral #1@}
\makeatother
\newcommand{\imth}{\hspace{1pt}\mathrm{i}\hspace{1pt}}

\newcommand{\mbz}{{\mathbb{Z}}}
\newcommand{\bea}{\begin{eqnarray}}
\newcommand{\eea}{\end{eqnarray}}
\newcommand{\bpm}{\begin{pmatrix}}
\newcommand{\epm}{\end{pmatrix}}
\newcommand{\bal}{\begin{aligned}}
\newcommand{\eal}{\end{aligned}}

\begin{document}
\title{Topological Crystalline Metal in Orthorhombic Perovskite Iridates}
\author{Yige Chen}
\affiliation{Department of Physics, University of Toronto, Ontario M5S 1A7 Canada}
\author{Yuan-Ming Lu}
\affiliation{Department of Physics, University of California, Berkeley, CA 94720}
\affiliation{Materials Sciences Division, Lawrence Berkeley National Laboratory, Berkeley, CA 94720}
\author{Hae-Young Kee}
\email{hykee@physics.utoronto.ca}
\affiliation{Department of Physics, University of Toronto, Ontario M5S 1A7 Canada}
\affiliation{Canadian Institute for Advanced Research, CIFAR Program in Quantum Materials, Toronto, ON M5G 1Z8, Canada}
\date{\today}

\begin{abstract}
Since topological insulators were theoretically predicted and experimentally observed in semiconductors with strong spin-orbit coupling, more and more attention has been drawn to topological materials which host exotic surface states. These surface excitations are stable against perturbations since they are protected by global or spatial/lattice symmetries. Succeeded in achieving various topological insulators, a tempting challenge now is to search for metallic materials with novel topological properties. Here we predict that orthorhombic perovskite iridates realize a new class of metals dubbed topological crystalline metals, which support zero-energy surface states protected by certain lattice symmetry. These surface states can be probed by photoemission and tunneling experiments. Furthermore, we show that by applying magnetic fields, the topological crystalline metal can be driven into other topological metallic phases, with different topological properties and surface states.
\end{abstract}
\maketitle

\section*{Introduction}

Recent discovery of topological insulators reveals a large class of new materials which, despite an insulating bulk, host robust metallic surface states\cite{Kane2005,Kane2005a,Bernevig2006a,Konig2007,Fu2007,Moore2007,Roy2009,Hsieh2008,Hsieh2009}. Unlike conventional ordered phases characterized by their symmetries, these quantum phases are featured by nontrivial topology of their band structures, and remarkably they harbor conducting surface states protected by global symmetries such as time reversal and charge conservation. More recently it was realized that certain insulators can support surface states protected by crystal symmetry, and they are named topological crystalline insulators\cite{Fu2011,Kargarian2013,Hsieh2014}. The rich topology of insulators in the presence of symmetries lead to a natural question: are there similar ``topological metals'' hosting protected surface states? After the proposal of Weyl semimetal\cite{Wan2011}, a large class of topological metals are classified\cite{Matsuura2013} which harbors surface flat bands protected by global symmetries such as charge conservation. However an experimental confirmation of these phases is still lacking\cite{Witczak-Krempa2014}.

In this work we propose that orthorhombic perovskite Iridates AIrO$_3$ where A is an alkaline earth metal, with strong spin-orbit coupling and Pbnm structure, can realize a new class of metal, dubbed topological crystalline metal (TCM). Topological properties of such a TCM phase include zero-energy surface states protected by the mirror reflection symmetry, and gapless helical modes located at the core of lattice dislocations. Photoemission and tunneling spectroscopy are natural experimental probes for these topological surface state. We further show how this TCM phase can be driven to metallic phases with different topology by applying magnetic fields, which breaks the mirror symmetry. All these results will be supported by topological classification in the framework of K-theory, as well as numerical calculations of topological invariants and surface/dislocation spectra, as presented in the Methods section.


\section*{Results}

\subsection{\bf Crystal structure and lattice symmetry in SrIrO$_3$}

Iridates have attracted much attention due to the strong spin-orbit coupling (SOC) in 5d-Iridium (Ir) and a variety
of crystal structures ranging from layered perovskites to pyrochlore lattices\cite{Kargarian2011,Yang2014}.
Despite structural differences, a common ingredient of these iridates with Ir$^{4+}$ is $J_{\rm eff}$=1/2
states governing low energy physics, resulted from a combination of strong SOC and crystal field splitting.
Among them, orthorhombic perovskite iridates AIrO$_3$ (where A is an alkaline earth metal) belongs
to Pbnm space group and can be tuned into a topological insulator (TI)\cite{Carter2012}.
%
%


A unit cell of AIrO$_3$ contains four Ir atoms as shown in Fig. 1(a), and there are three types of symmetry plane: b-glide, n-glide and mirror plane perpendicular to $\hat{\bf c}$-axis. Each of them can be assigned to the symmetry operators: $\Pi_b$, $\Pi_n$ and $\Pi_m$ as listed in the Methods section.
Introducing three Pauli matrices corresponding to the in-plane sublattice (${\boldsymbol\tau}$), layer (${\boldsymbol\nu}$), and
pseudospin $J_{\rm eff}$=1/2 (${\boldsymbol\sigma}$),
the tight-binding Hamiltonian $H({\bf k})$\cite{Carter2012} has a relatively simple form shown in Eq. (\ref{eq:m3}).\\

The band structure of tight-binding Hamiltonian $H({\bf k})$ exhibits a ring-shaped one-dimensional (1D) Fermi surface (FS) close to the Fermi level as shown in Fig. 1(b), which we call the nodal ring, as the energy dispersion is linear in two perpendicular directions. This nodal ring FS was confirmed by first-principle ab-initio calculations\cite{Carter2012,Zeb2012},  and it remains intact in the presence of Hubbard $U$ up to 2.5 eV. Therefore, the semimetallic character of SrIrO$_3$ with $U$ around 2 eV is consistent with previous experimental results~\cite{Cao2007, Moon2008}.
It was further shown that the size of nodal ring is determined by rotation and tilting angles of the oxygen octahedra
around each Ir atom. In the case of SrIrO$_3$ which has a minimal distortion of octahedra, the nodal ring is centered around the U$\equiv(k_a=0,k_b=\frac{\pi}{a},k_c=\frac{\pi}{c})$ point, and extends in two-dimensional (2D) U-R-S-X plane (perpendicular to ${\hat {\bf b}}$-axis) in three-dimensional (3D)
Brillouin zone (BZ) as shown in Fig. 1(b).\\


\subsection{\bf Zero-energy surface states and dislocation helical modes}

Here we show that the nodal ring FS exhibits non-trivial topology that leads to localized surface zero modes protected by the mirror symmetry, thus coined topological crystalline metal (TCM).
%
To demonstrate the existence of zero-energy surface states, the band structure calculation for the $\hat{ {\bf a}}\hat{{\bf c}}$-side plane was carried out
with open boundary i.e. $[1\bar10]$ surface perpendicular to ${\hat{\bf b}}$-axis.
%
%
The energy dispersion at $k_c=\frac\pi c$, as displayed in Fig. 1(c), reveals a dispersionless zero-energy flat band
marked by red color for all $k_a$ on $[1\bar10]$ surface of the sample. On the other hand, the slab spectrum at $k_a=0$ shows the surface states are gapped
except at $k_c =\frac\pi c$ as shown in Fig. 1(d).
%
%
Similar calculations for $\hat{\bf b}\hat{\bf c}$-side plane show that $[110]$ surface (perpendicular to ${\hat {\bf a}}$-axis) also supports localized surface zero-modes at $k_c=\frac\pi c$, while $[010]$ surface (perpendicular to $\hat{\bf y}$-axis) does not harbor any zero-energy states. As elaborated in the Methods section, these surface states manifest a mirror-symmetry-protected weak index labeled by a vector\cite{Ran2010} ${\bf M}={\hat {\bf a}}+{\hat {\bf b}}~//~\hat{\bf y}$, where ${\hat {\bf a}}$ and ${\hat {\bf b}}$ are Bravais lattice primitive vectors. A direct consequence of this weak index is the existence of $k_c=\frac\pi c$ zero-energy states for \emph{any} side surface, except for $[010]$ surface perpendicular to vector ${\bf M}$.

Another consequence of weak index ${\bf M}$ is the existence of pairs of counter-propagating zero modes (``helical modes'') localized in a dislocation line, which respects mirror symmetry $\Pi_m$. The number of zero mode pairs $N_0$ in each dislocation line is determined by its Burgers vector ${\bf B}$ by $N_0={\bf B}\cdot{\bf M}/2\pi$.\cite{Ran2009} We've performed numerical calculations which demonstrate a pair of gapless helical modes in a dislocation line along $\hat{\bf c}$-axis with Burgers vector ${\bf B}=\pm{\hat{\bf a}}$. Detailed results are presented in the Methods section.\\

\subsection{\bf Classification and topological invariants}

To understand the topological nature of the zero-energy surface states,
let us first clarify the symmetry of tight-binding model $H({\bf k})$. It turns out in the mirror-reflection-symmetric $k_c=\frac\pi c$ plane, $H({\bf k})$ has an \emph{emergent} chiral symmetry, i.e.
$\{H({\bf k}) , {\mathcal C}\} =0$ where ${\mathcal C} =  \sigma_z \nu_x \tau_z$. Here $\boldsymbol\sigma$,$\boldsymbol\nu$ and $\boldsymbol\tau$ represent $J_{\rm eff}=\frac{1}{2}$ pesudo-spin space, inter-plane and in-plane sublattice, respectively. A chiral symmetry can be understood as the combination of time reversal symmetry and certain particle-hole symmetry\cite{Schnyder2008}, hence switching the sign of Hamiltonian. The presence of chiral symmetry $\mathcal{C}$ enforces the energy spectrum of $H({\bf k})$ to be symmetrical with respect to the zero energy. It is known that chiral symmetry can protect zero-energy surface modes\cite{Heeger1988,Schnyder2011}. On the other hand, various crystal symmetries such as mirror reflection $\Pi_m$, bring extra non-trivial topological properties into the system we studied.  Starting from the $k_c=\frac\pi c$ plane with mirror reflection $\Pi_m=\sigma_z\nu_x$ and chiral symmetry $\mathcal{C}$, we can classify possible surface flat bands in the mathematical framework of K-theory\cite{Kitaev2009}, as summarized in Table \ref{tab:classification:[xy0]} (see Methods section for details).

In particular with both mirror and chiral symmetries, the classification is $\mbz\times\mbz$ characterized by a pair of integer topological invariants $(W^+,W^-)$. Since the Hilbert space can be decomposed into two subspaces with different mirror eigenvalues $\Pi_m=\pm1$, in each subspace we can obtain a 1D winding number\cite{Matsuura2013}
\begin{equation}
W^{\pm}_{[lmn]}({\bf k}_{\|})=\frac{1}{2\pi \imth} \int d k_{\bot} {\rm Tr}[\mathcal{C} h_{\pm}^{-1}({\bf k})\frac{\partial}{\partial k_{\bot}} h_{\pm}({\bf k})]\; ,
\label{eq:w1}
\end{equation}
where $k_{\bot}$ is the crystal momentum along $[lmn]$ direction, and $h_\pm({\bf k})=H^{\pm}_{{\bf k}_{\|}}(k_{\bot})$ is the 1D Hamiltonian parametrized by $[lmn]$ surface momentum ${\bf k}_{\|}$ in $\Pi_m=\pm1$ subspace. For both $\hat{\bf b}\hat{\bf c}$-plane ($[110]$
surface) and $\hat{\bf a}\hat{\bf c}$-plane ($[1\bar10]$ surface) we have $(W^+,W^-)=(1,-1)$.
These quantized winding numbers correspond to a pair of zero modes for each surface momentum with $k_c=\frac{\pi}{c}$, and they cannot hybridize due to opposite mirror eigenvalues. Meanwhile for $[010]$ surface both $W^{\pm}$ vanish, indicating a weak index\cite{Ran2010} ${\bf M}=\hat{\bf a}+\hat{\bf b}~//~\hat{\bf y}$. Intuitively the system can be considered as a stacked array of $\hat{\bf x}\hat{\bf z}$-planes, each plane (perpendicular to $\hat{\bf y}$-axis) with a pair of mirror-protected zero-energy edge modes at $k_c=\frac{\pi}{c}$.

Once the mirror symmetry $\Pi_m$ is broken, the classification becomes $\mbz$ characterized by one integer topological invariant: the total winding number $W\equiv W^++W^-$ associated with 1D TI in symmetry class AIII\cite{Schnyder2008,Kitaev2009}. This total winding number vanishes for all surface momenta at $k_c=\frac\pi c$ though.
\newline
%



\subsection{\bf Topological metal/semimetal induced by magnetic fields} 

Time-reversal (TR) breaking perturbations like a magnetic field can drive the system from the TCM phase to other metallic phases with different topological properties. In the presence of Zeeman coupling $\mu_B{\bf h}\cdot{\boldsymbol\sigma}$ introduced by magnetic field ${\bf h}$, clearly the chiral symmetry $\mathcal{C}$ is still preserved as long as the field is in $\hat{\bf x}\hat{\bf y}$-plane (${\bf h}\perp\hat{\bf z}$). Magnetic field parallels to $\hat{\bf z}$ axis breaks chiral symmetry, which gaps the nodal ring, making the system trivial.  Meanwhile mirror symmetry $\Pi_m$ will be broken unless the magnetic field is along $\hat{\bf z}$-direction. Thus, our focus below will be magnetic field in the $\hat{\bf x}\hat{\bf y}$-plane. The FS topology and associated surface states with a magnetic field along different directions are listed in Table~\ref{tab:dp}.



\subsection{\bf (i) Chiral topological metal protected by chiral symmetry} 

Due to TR and inversion symmetry, the nodal ring in TCM always has 2-fold Kramers degeneracy.
After applying $[1\bar{1}0]$ direction magnetic field ${\bf h}//\hat{\bf b}$, the doubly-degenerate nodal ring in Fig. 1(b) splits into two rings in Fig. 2(a) shifted along $\hat{\bf c}(\hat{\bf z})$-axis on U-R-S-X plane.
%
%
Though the mirror symmetry is broken by the $[1\bar10]$ magnetic field, chiral symmetry $\mathcal{C}$ is still preserved. Therefore the topological properties of the two nodal rings are captured by an integer winding number $W$ in the symmetry class AIII\cite{Matsuura2013}. Consider $[1\bar{1}0]$ surface for instance, depending on the surface momentum ${\bf k}_{\|}$, the winding number $W_{{\bf k}_{\|}}$ is plotted in Fig. 2(b). It vanishes in the region where two nodal rings (blue and red) overlap, but becomes $\pm1$ in other regions within two nodal rings.

The energy spectra on a slab geometry with open $[1\bar{1}0]$ surfaces are displayed in Fig. 2(c) and 2(d), plotted as a function of $k_a$ with $k_c=\frac{\pi}{c}$ and $k_c=\frac{\pi}{c}-\delta$, respectively. There is no zero modes at $k_c=\frac\pi c$, corresponding to trivial winding number. Meanwhile zero-energy flat bands highlighted by red color in Fig. 2(d) exist inside the nodal rings. It confirms the non-trivial topology of the bulk nodal rings with quantized winding number $W_{[1\bar{1}0]}=\pm1$ shown in Fig. 2(b).
%
It turns out this chiral topological metal supports localized flat band protected by chiral symmetry on any surface, as long as its normal vector $\hat{\bf n}$ is not perpendicular to $\hat{\bf b}$-axis. Meanwhile the two nodal rings are stable against any perturbations preserving chiral symmetry, since the winding number changes when we cross each nodal ring\cite{Matsuura2013}.


\subsection{\bf (ii) Weyl semimetal} 

Once we apply a magnetic field along $[110]$ direction (or $\hat{\bf a}$-axis), both mirror ($\Pi_m$) and n-glide ($\Pi_n$) symmetries are broken. Consequently the nodal ring is replaced by a pair of 3D Dirac nodes, appearing at momenta $(\pm k_{0},\frac{\pi}{a},\frac{\pi}{c})$ along the path R$\rightarrow$U$\rightarrow$R  BZ line. However these Dirac nodes are not symmetry protected, since a sublattice potential $m\nu_z$ alternating by layers would further split each Dirac point into a pair of Weyl nodes. And this $m\nu_z$ term has the same symmetry as a Zeeman field $h_{[110]}$ along $\hat{\bf a}$-axis.

In the presence of both magnetic field $h_{[110]}$ and layer-alternating potential $m\nu_z$, the system still preserves chiral symmetry $\mathcal{C}$, b-glide $\Pi_b$ and inversion symmetry. Two pairs of Weyl nodes emerge at $\pm{\bf k}_1=(\pm  k_{1},\frac{\pi}{a},\frac{\pi}{c})$ and $\pm{\bf k}_2=(\pm k_{2},\frac{\pi}{a},\frac{\pi}{c})$ along R$\rightarrow$U$\rightarrow$ R line in Fig. 3(a). The low-energy Hamiltonian around ${\bf k}_{1}$ has linear dispersion along all ${\bf k}$ directions
\begin{align}
&H_{\rm eff}(\delta {\bf k}) = {\bf p}(\delta {\bf k})\cdot {\boldsymbol\sigma},\label{eq:w1}\\
&\notag {\bf p}(\delta {\bf k})\equiv (p_x,p_y,p_z)=\notag\Big(A_1 \delta k_b-A_2 \delta k_c,B_1 \delta k_c,D_1 \delta k_a\Big),\label{eq:weyl}
\end{align}
with $\delta {\bf k} \equiv {\bf k} - {\bf k}_1$. Various coefficients can be expressed in terms of the tight-binding hopping parameters (see Methods). 

There exists a ``jump'' for the Chern number $C$ for all occupied bands from $C=0$ to $C=1$ when $k_a$ crosses $k_1$, and similarly an opposite ``jump'' from $C=1$ to $C=0$ after $k_a$ passes $k_2$. The different signs of jumps in Chern number indicate the Weyl fermions at $\pm{\bf k}_1$ and $\pm{\bf k}_2$ has opposite topological charge $+1$ (blue) and $-1$ (red) respectively, as shown in Fig. 3(a).

The surface states on $[1\bar{1}0]$ surface at $k_a=\frac{0.7}{a}$, between the two Weyl nodes with opposite chirality, are plotted along $k_c$ in Fig. 3(b). There is a single dispersing zero mode colored with red in Fig. 3(b) which is localized on each surface of the sample.  A series of one-way-dispersing zero modes for all surface momenta between ${\bf k}_1$ and ${\bf k}_2$  form a ``chiral Fermi arc''\cite{Wan2011,Yang2011} on $[1\bar{1}0]$ surface, as shown by the green lines in Fig. 3(a).
\newline

\section*{Discussion}

The existence of surface zero modes in our TCM phase originates from the chiral and mirror reflection symmetry of AIrO$_3$ with Pbnm structure. Any side surface other than $[010]$-plane should exhibit robust zero-energy surface states independent of the details.
In a generic band structure of SrIrO$_3$, this nodal ring does not
sit exactly at the Fermi level $E_{\rm F}$, but slightly below $E_{\rm F}$ (unless SOC is stronger than an atomic SOC used in the first-principle calculation), and
a hole-like pocket FS occurs around $\Gamma$ point in Fig. 1(b)~\cite{Zeb2012}. In other words, the nodal ring occurs around $k_c=\frac\pi c$, while the small bulk Fermi pocket is located near $k_c=0$.  Therefore,the zero-energy surface modes are well separated from the bulk FS pockets in momentum, and a momentum resolved probe is required to detect the surface states. Angle-resolved photoemission spectroscopy (ARPES) would be the best tool to observe the momentum-resolved surface states shown in Fig. 1(c) and (d) on a side-plane of AIrO$_3$. Notice that, ARPES has successfully detected the topological surface states in Dirac semimetal material\cite{Yi2014}. Due to the presence of extremely small orbital overlap amplitudes between further Ir sites, the surface states acquire a slight dispersion. However, as we emphasized above, the mirror symmetry is a crucial ingredient to support such non-trivial surface states detectable by ARPES, despite the ``weak breaking'' of chiral symmetry.

These surface states also contribute finite surface density of states (SDOS) near zero energy. In contrast, the semimetallic bulk band contribution to SDOS vanishes around zero energy due to the presence of bulk nodal ring. Therefore, protected surface modes can be detected as a zero bias hump (finite SDOS) in the dI/dV-curve of scanning tunneling microscopy (STM). However, in real materials, it will be difficult to separate the contribution of the surface states to SDOS from the bulk part. On the other hand, there exists protected propagating fermion modes in dislocation lines\cite{Ran2009} that preserves mirror symmetry. One advantage is that these topological helical modes are protected by mirror and chiral symmetries and hence won't be destroyed by hybridization with bulk gapless excitations. In particular counter-propagating gapless fermions show up in pairs in the dislocation core, and the number of gapless fermion pairs is given by ${\bf M}\cdot{\bf B}/2\pi$ where ${\bf B}$ is the Burgers vector of dislocation as shown
in the Methods section. Unlike other gapless and dispersive bulk excitations that are extended in space, these helical fermion zero-modes are localized near the dislocation core, which in hence can be detected by STM.

Since a bulk sample of AIrO$_3$ such as SrIrO$_3$ requires high pressure to achieve the Pbnm crystal structure\cite{Longo1971,Zhao2008},
it is desirable to grow a film of AIrO$_3$. Recently superlattices of atomically thin slices of SrIrO$_3$ was made by pulsed laser deposition along [001] plane\cite{Matsuno2014}. This work is a first step towards possible topological phases in iridates. Furthermore, a successful growth of film along [111] plane was also report\cite{Takagi14}. Thus growing a film of AIrO3 along [110] (or [1-10]) is plausible. To confirm the proposed TCM, an ARPES  study should be performed on a film of AIrO3 grown along [110] (or [1-10]), where the mirror symmetry of Pbnm structure is kept. This analysis should reveal a flat surface band near $k_c$= $\frac{\pi}{c}$ below $E_{\rm F}$.\\

\section*{Methods}

\subsection{\bf Symmetry operators and tight-binding Hamiltonian}

The tight-binding Hamiltonian is defined in the basis of 8-component spinor $\psi$, organized as\cite{Carter2012}
\beq
\psi\equiv(c_{B\uparrow},c_{R\uparrow},c_{Y\uparrow},
c_{G\uparrow},c_{B\downarrow},c_{R\downarrow},c_{Y\downarrow},c_{G\downarrow})^T\; ,
\eeq
where $B,R,Y,G$ correspond to 4 sublattices. We define Pauli matrices ${\boldsymbol\tau}$ and ${\boldsymbol\nu}$ in terms of following sublattice rotations
\bea
& \notag B\overset{\tau_x}\longleftrightarrow R,~~Y\overset{\tau_x}\longleftrightarrow G;\\
& \label{eq:m2} B\overset{\nu_x}\longleftrightarrow Y,~~R\overset{\nu_x}\longleftrightarrow G.
\eea

The full space group Pbnm is generated by (see FIG. 2 in Ref. 18) translations $T_{x,y,z}$ (three Bravais primitive vectors correspond to $T_a\equiv T_xT_y,~T_b\equiv T_x^{-1}T_y$ and $T_c=T_z^2$) and the following three generators $\{\Pi_b,\Pi_n,\Pi_m\}$
\bea
&\notag\psi_{(k_a,k_b,k_c)}\overset{\Pi_n}\longrightarrow e^{\imth\frac{k_a+k_c}2}\frac{\imth(\sigma_x-\sigma_y)}{\sqrt2}\nu_x\tau_x \cdot e^{\imth \frac{k_c \nu_z+k_a \tau_z}2} \psi_{(k_a,-k_b,k_c)},\\
&\notag\psi_{(k_a,k_b,k_c)}\overset{\Pi_b}\longrightarrow e^{\imth\frac{k_b}2}\frac{\imth(\sigma_x+\sigma_y)}{\sqrt2}\tau_x \cdot e^{\imth \frac{k_b \tau_z}2}\psi_{(-k_a,k_b,k_c)},\\
&\label{symmetry matrix}\psi_{(k_a,k_b,k_c)}\overset{\Pi_m}\longrightarrow\imth\sigma_z\nu_x \psi_{(k_a,k_b,-k_c)},\\
&\notag\psi_{\bf k}\overset{\bsi}\longrightarrow\psi_{-{\bf k}}.
\eea
where $\Pi_b$ (glide plane $\perp\hat{\bf a}\equiv\hat{\bf x}+\hat{\bf y}$) and $\Pi_n$ (glide plane $\perp\hat{\bf b}\equiv\hat{\bf y}-\hat{\bf x}$) represents two glide symmetries while $\Pi_m$ is the mirror reflection and ${\boldsymbol\sigma}$ denotes the pseudo-spin subspace. $\bsi=\Pi_b\Pi_n\Pi_m$ represents the inversion symmetry.


The tight-binding Hamiltonian of SrIrO$_3$ has the following form~\cite{Carter2012}
\bea
H_{\bf k}&=&{\rm Re}(\epsilon^{\rm p}_{\bf k}) \tau_x + {\rm Im}(\epsilon^{\rm p}_{\bf k}) \sigma_z\tau_y +\epsilon^{\rm z}_{\bf k} \nu_x\nonumber\\
&+&{\rm Re}(\epsilon^{\rm d}_{\bf k}) \nu_x\tau_x + {\rm Im}(\epsilon^{\rm d}_{\bf k})\nu_y\tau_y+{\rm Re}(\epsilon^{\rm po}_{\bf k})\sigma_y\nu_z\tau_y\nonumber\\
&+& {\rm Im}(\epsilon^{\rm po}_{\bf k})\sigma_x\nu_y\tau_z
+{\rm Re}(\epsilon^{\rm do}_{\bf k})\sigma_y\nu_x\tau_y +{\rm Im}(\epsilon^{\rm do}_{\bf k})\sigma_x \nu_x\tau_y\nonumber\\
&+&{\rm Re}(\epsilon^{\rm d1}_{\bf k})\sigma_y\nu_y\tau_x + {\rm Im}(\epsilon^{\rm d1}_{\bf k})\sigma_x\nu_y\tau_x\; .
\label{eq:m3}
\eea
Here various coefficient functions have been defined in Ref. 18 with additional term $\epsilon^{\rm d1}_{\bf k}$
\begin{equation}
\epsilon^{\rm d1}_{\bf k}=t_{\rm d1}(\cos(k_y)+ \imth \cos(k_x))\cos(k_z)\; ,
\label{eq:m4}
\end{equation}
where $t_{\rm 1d}$ is the inter-layer next nearest neighbor hopping due to non-vanishing rotation and tilting in local oxygen octahedra. The crystal momentum $(k_a,k_b,k_c)$ relates to $(k_x,k_y,k_z)$ simply by
\bea
k_a=\frac{k_x+k_y}a, \quad k_b=\frac{k_y-k_x}a, \quad k_c=\frac{2k_z}c\label{momentum}\; .
\label{eq:m5}
\eea
 The basis $\psi_{(k_a,k_b,k_c)}$ we use here is different with the basis $\psi_{(k_x,k_y,k_z)}$ in Ref. 18 by an unitary rotation $U_{\bf k}$
\begin{eqnarray}
\psi_{(k_a,k_b,k_c)}\equiv U_{\bf k}^{\dagger} \psi_{(k_x,k_y,k_z)}\; , \quad U_{\bf k}=e^{\imth \frac{k_z}2 \nu_z}e^{\imth \frac{k_x}2 \tau_z}\; .
\label{eq:m6}
\end{eqnarray}
In this basis $\psi_{(k_a,k_b,k_c)}$, the tight-binding Hamiltonian is related to $H_{\bf k}$ by the unitary transformation $U_{\bf k}$ in Eq.~\ref{eq:m3}:
\bea
H({\bf k})\equiv H(k_a,k_b,k_c)=U^{\dagger}_{\bf k}H_{\bf k} U_{\bf k}\; .
\label{eq:m7}
\eea


Clearly in $k_c=\frac\pi c$(or $k_z=\frac{\pi}2$) plane the nonvanishing terms in $H_{\bf k}$, as in the basis $\psi_{(k_x,k_y,k_z)}$, contain only Pauli matrices $(\tau_x,\sigma_z\tau_y),~\nu_y\tau_y$ and $\sigma_{x,y}\nu_{x,z}\tau_y$. All these Pauli matrices anticommute with
\bea
\mathcal{C}=\sigma_z\nu_y\tau_z\; ,
\label{eq:m8}
\eea
\ie they all have chiral symmetry $\mathcal{C}$. However, in $\psi_{(k_a,k_b,k_c)}$ representation, the chiral symmetry $\mathcal{C}$ written as
\bea
\mathcal{C}=\sigma_z\nu_x\tau_z\label{chiral symmetry}\; .
\eea

\subsection{\bf K-theory classification procedure and topological invariants}

To understand surface states on $[1\bar10]$ surface for instance, let's focus on a 1d system $H_{\bf k}$ in momentum space parametrized by fixed momentum $k_a\in[-\frac{\pi}a,\frac{\pi}a)$ and $k_c=\frac{\pi}c$. Such a 1d system will have mirror reflection $\Pi_m=\imth\sigma_z\nu_x$ in (\ref{symmetry matrix}) as well as chiral symmetry  $\mathcal{C}=\sigma_z\nu_x\tau_z$ in (\ref{chiral symmetry}). We'll classify such a gapped 1d system (since the bulk gap only closes at two points in $k_c=\frac{\pi}c$ plane) to see whether it has nontrivial topology, which may protect gapless surface states.

The classification of a gapped system can be understood from classifying possible symmetry-allowed mass matrices for a Dirac Hamiltonian\cite{Kitaev2009,Wen2012}. The mathematical framework of K-theory applies to both global symmetry and certain spatial (crystal) symmetry\cite{Morimoto2013,Lu2014a}. In 1d such a Dirac Hamiltonian can be written as
\bea
H_{Dirac}^{1d}=k_a\gamma_1+m\gamma_0\;,
\eea
with chiral symmetry $\mathcal{C}$
\bea
\{\mathcal{C},\gamma_0\}=\{\mathcal{C},\gamma_1\}=0.
\eea
mirror reflection $\Pi_m$
\bea
[\Pi_m,\gamma_0]=[\Pi_m,\gamma_1]=0.
\eea
and $U(1)$ charge conservation $Q$
\bea
[Q,\gamma_0]=[Q,\gamma_1]=0.
\eea
Any two symmetry generators among $(\mathcal{C},\Pi_m,Q)$ commute with each other. Mathematically the classification problem correspond to the following question: given Dirac matrix $\gamma_1$ and symmetry matrices $(\mathcal{C},\Pi_m,Q)$, what is the \emph{classifying space} $\mathcal{S}$ of mass matrix $\gamma_0$? In particular since each \emph{disconnected piece} in classifying space $\mathcal{S}$ correspond to one gapped 1d phase, how many disconnected pieces does $\mathcal{S}$ contain? Hence the classification of gapped 1d phases with these symmetries is given by the zeroth homotopy $\pi_0(\mathcal{S})$.

In the K-theory classification, if there are generators which commute with all Dirac matrices and other symmetry generators, such as $U(1)$ charge symmetry generator $Q$ here satisfying $Q^2=(-1)^F$, then we say the gapped system belong to a \emph{complex} class. Otherwise it belongs to a \emph{real} class. Clearly our case belongs to the complex class because both $Q$ and $M$ commute with any other matrices.

For the $k_c=\pi/c$ plane on a generic $[xy0]$ side surface parallel to $\hat{c}$-axis, the full symmetry group is generated by $(\mathcal{C},\Pi_m,Q)$ as mentioned earlier. Note that these 3 symmetry generators all commute with each other while $Q^2=(-1)^F$. Together with Dirac matrix $\gamma_1$ they form a complex Clifford algebra $Cl_2\times Cl_2$:
\bea
\{\gamma_1,\mathcal{C}\}\times\Pi_m\times Q
\eea
where the generators inside the parenthesis anti-commute with each other and commute with everything outside the parenthesis. The reason we have $Cl_2\times Cl_2$ is because we can block-diagonalize the $k_c=\pi$ tight-binding Hamiltonian w.r.t. their $\Pi_m$ (mirror reflection) eigenvalue $\pm1$, and in each subspace the Clifford algebra is $Cl_2$ (two generator in the parenthesis). Now when we add the mass matrix $\gamma_0$ the complex Clifford algebra is extended to $Cl_3\times Cl_3$ generated by
\bea
\{\gamma_1,\mathcal{C},\gamma_0\}\times\Pi_m\times Q
\eea
Therefore the classifying space of mass matrix $\gamma_0$ is determined by the extension problem of Clifford algebra $Cl_2\times Cl_2\rightarrow Cl_3\times Cl_3$ and we label such a classifying space as $\mathcal{S}=C_2\times C_2$. The classification of gapped 1d phases with $(\mathcal{C},\Pi_m,Q)$ symmetries is hence given by
\bea
\pi_0(C_2\times C_2)=\pi_0(C_2)\times\pi_0(C_2)=\mbz\times\mbz
\eea
There are two integer-valued topological invariants $(W^+,W^-)$ which are 1d winding numbers\cite{Schnyder2011} obtained in block-diagonalized subspace with $\Pi_m=\sigma_z\nu_y=\pm1$.

Now let's start to break $\Pi_m$ and $\mathcal{C}$ symmetries. When we break mirror reflection $\Pi_m$ (but keep $\mathcal{C}$) the Clifford algebra extension problem becomes $Cl_2\rightarrow Cl_3$
\bea
\{\gamma_1,\mathcal{C}\}\times Q\rightarrow\{\gamma_1,\mathcal{C},\gamma_0\}\times Q
\eea
and hence the classification is $\pi_0(C_2)=\mbz$. The topological invariant is total winding number $W=W^++W^-$.

If we break $\mathcal{C}$ but keep $\Pi_m$, the extension problem is again $(Cl_1)^2\rightarrow (Cl_2)^2$
\bea
\{\gamma_1\}\times\Pi_m\times Q\rightarrow\{\gamma_1,\gamma_0\}\times\Pi_m\times Q
\eea
and classification is trivial since
\bea
\pi_0(C_1\times C_1)=\pi_0(C_1)\times\pi_0(C_1)=0\times0=0.
\eea

If we break both $\mathcal{C}$ and $\Pi_m$ symmetries, our extension problem is $Cl_1\rightarrow Cl_1$
\bea
\{\gamma_1\}\times Q\rightarrow\{\gamma_1,\gamma_0\}\times Q
\eea
which leads to a trivial classification $\pi_0(C_1)=0$. Therefore we obtain the results in Table I.

\subsection{\bf Dislocation Spectrum}

One implication of weak index ${\bf M}$ is the existence of pairs of counter-propagating zero modes localized in a dislocation line respecting mirror symmetry $\Pi_m$. The number of zero mode pairs in each dislocation line is determined by its Burgers vector ${\bf B}$: number of helical modes$={\bf B}\cdot{\bf M}/2\pi$.\cite{Ran2009}

%
To check such zero modes due to dislocation line,
we consider a pair edge dislocations along $\hat{\bf c}$-axis, with a pair of Burgers vectors: (${\bf B}=\hat{\bf a}$, ${\bf B}=-\hat{\bf a}$) perpendicular to the dislocation line. The type of dislocation core is illustrated in Fig. 4(a).  We consider now a 3D box with periodic boundary condition in all ($\hat{\bf a},\hat{\bf b},\hat{\bf c}$) directions (a 3D-torus) so that the system does not have an open surface. Consider every unit cell (with 4 sublattices, 2 pseudo-spin species per site) as a lattice `site' in Fig. 4(a), then the system has hopping terms between nearest neighbor `sites' following the tight-binding Hamiltonian in Eq.~(\ref{eq:m3}). $\hat{\bf c}$ direction is still translational invariant and $k_z$ remains a good quantum number, but in $\hat{\bf a}\hat{\bf b}$-plane translation symmetry is broken by the dislocation.
The dislocation spectrum is obtained when there are 39 sites along $\hat{\bf a}$-axis and 15 sites along $\hat{\bf b}$-axis. The location of dislocation with Burgers vector ${\bf B}=\hat{\bf a}$ is at  $(4,6)$ which means 4th site along $\hat{\bf a}$-axis and 6th site along $\hat{\bf b}$-axis. And the position of dislocation with Burgers vector ${\bf B}=-\hat{\bf a}$ is at $(24,12)$. The dislocation spectrum has displayed in Fig. 4(b) highlighted by red color.
It shows two pairs of gapless helical modes localized on each dislocation cores.

\subsection{\bf Effective Hamiltonian of Weyl fermion}

By projecting the states around the Weyl node, the two-band Hamiltonian with linear Weyl fermion form can be obtained after adding $m \nu_z$ which breaks mirror symmetry and magnetic field $h_{[110]}(\sigma_x+\sigma_y)$ which breaks TR symmetry to $H({\bf k})$ in Eq.~(\ref{eq:m3}). The followings are the coefficients for the effective Hamiltonian describes the Weyl fermion at ${\bf k}_1$ of Eq. (2) presented in the maintext:
\begin{eqnarray}
A_1&=&\frac{2 t_{\rm p} t_{\rm d}^{\rm o}\sin^2(k_1)}{(t_{\rm 2p}^{\rm o}-t_{\rm 1p}^{\rm o})\cos(k_1)}\; ,\nonumber\\ A_2&=&\frac{(\sqrt{2} h -m)t_{\rm z}^{\rm o}}{\sqrt{2}(t_{\rm 2p}^{\rm o}-t_{\rm 1p}^{\rm o})\cos(k_1)}\; ,\nonumber\\ B_1&=&\frac{t_{\rm d1} (m-\sqrt{2}h)}{\sqrt{2}(t_{\rm 2p}^{\rm o}-t_{\rm 1p}^{\rm o})}\; ,\nonumber\\
D_1&=&\frac{[(t_{\rm 2p}^{\rm o}-t_{\rm 1p}^{\rm o})^2-(t_d^o)^2]\sin(k_1)}{t_{\rm 2p}^{\rm o}-t_{\rm 1p}^{\rm o}}\; ,
\label{eq:weyl1}
\end{eqnarray}
where $t_{\rm p},t_{\rm 2p}^{\rm o},t_{\rm 1p}^{\rm o},t_{\rm d}^{\rm o},m,t_{\rm d1},h\equiv h_{[110]}$ are the coefficients in tight-binding Hamiltonian. The Chern number $C$ for all occupied bands as a function of crystal momentum along ${\rm U} \rightarrow {\rm R}$ (or $k_a$) is shown in Fig. 4(c).\\





\paragraph{\bf Acknowledgments}
This work is supported by Natural Science and Engineering Research Council of Canada (NSERC), Center for Quantum Materials at the University of Toronto (YC and HYK),
and Office of BES, Materials Sciences Division of the U.S. DOE under contract No. DE-AC02-05CH11231 (YML).
HYK thanks S. Ryu for informing topology of gapless superconductors in Ref. 36.
YML and HYK acknowledge the hospitality of the Aspen Cetner for Physics supported by National Science Foundation Grant No. PHYS-1066293,
where a part of this work was carried out.


\section*{Reference}

\begin{thebibliography}{30}
\expandafter\ifx\csname url\endcsname\relax
  \def\url#1{\texttt{#1}}\fi
\expandafter\ifx\csname urlprefix\endcsname\relax\def\urlprefix{URL }\fi
\providecommand{\bibinfo}[2]{#2}
\providecommand{\eprint}[2][]{\url{#2}}

\bibitem{Kane2005}
\bibinfo{author}{Kane, C.~L.} \& \bibinfo{author}{Mele, E.~J.}
\newblock \bibinfo{title}{Quantum spin hall effect in graphene}.
\newblock \emph{\bibinfo{journal}{Phys. Rev. Lett.}}
  \textbf{\bibinfo{volume}{95}}, \bibinfo{pages}{226801}
  (\bibinfo{year}{2005}).
\newblock \urlprefix\url{http://link.aps.org/doi/10.1103/PhysRevLett.95.226801}.

\bibitem{Kane2005a}
\bibinfo{author}{Kane, C.~L.} \& \bibinfo{author}{Mele, E.~J.}
\newblock \bibinfo{title}{$Z_{2}$ topological order and the quantum spin hall
  effect}.
\newblock \emph{\bibinfo{journal}{Phys. Rev. Lett.}}
  \textbf{\bibinfo{volume}{95}}, \bibinfo{pages}{146802}
  (\bibinfo{year}{2005}).
\newblock \urlprefix\url{http://link.aps.org/doi/10.1103/PhysRevLett.95.146802}

\bibitem{Bernevig2006a}
\bibinfo{author}{Bernevig, B.~A.}, \bibinfo{author}{Hughes, T.~L.} \&
  \bibinfo{author}{Zhang, S.-C.}
\newblock \bibinfo{title}{Quantum spin hall effect and topological phase
  transition in hgte quantum wells}.
\newblock \emph{\bibinfo{journal}{Science}} \textbf{\bibinfo{volume}{314}},
  \bibinfo{pages}{1757--1761} (\bibinfo{year}{2006}).
\newblock \urlprefix\url{http://www.sciencemag.org/content/314/5806/1757.abstract}.

\bibitem{Konig2007}
\bibinfo{author}{Konig, M.} \emph{et~al.}
\newblock \bibinfo{title}{Quantum spin hall insulator state in HgTe quantum
  wells}.
\newblock \emph{\bibinfo{journal}{Science}} \textbf{\bibinfo{volume}{318}},
  \bibinfo{pages}{766--770} (\bibinfo{year}{2007}).
\newblock \urlprefix\url{http://www.sciencemag.org/content/318/5851/766.abstract}.

\bibitem{Fu2007}
\bibinfo{author}{Fu, L.}, \bibinfo{author}{Kane, C.~L.} \&
  \bibinfo{author}{Mele, E.~J.}
\newblock \bibinfo{title}{Topological insulators in three dimensions}.
\newblock \emph{\bibinfo{journal}{Phys. Rev. Lett.}}
  \textbf{\bibinfo{volume}{98}}, \bibinfo{pages}{106803}
  (\bibinfo{year}{2007}).
\newblock \urlprefix\url{http://link.aps.org/doi/10.1103/PhysRevLett.98.106803}.

\bibitem{Moore2007}
\bibinfo{author}{Moore, J.~E.} \& \bibinfo{author}{Balents, L.}
\newblock \bibinfo{title}{Topological invariants of time-reversal-invariant
  band structures}.
\newblock \emph{\bibinfo{journal}{Phys. Rev. B}} \textbf{\bibinfo{volume}{75}},
  \bibinfo{pages}{121306} (\bibinfo{year}{2007}).
\newblock \urlprefix\url{http://link.aps.org/doi/10.1103/PhysRevB.75.121306}.

\bibitem{Roy2009}
\bibinfo{author}{Roy, R.}
\newblock \bibinfo{title}{Topological phases and the quantum spin hall effect
  in three dimensions}.
\newblock \emph{\bibinfo{journal}{Phys. Rev. B}} \textbf{\bibinfo{volume}{79}},
  \bibinfo{pages}{195322} (\bibinfo{year}{2009}).
\newblock \urlprefix\url{http://link.aps.org/doi/10.1103/PhysRevB.79.195322}.

\bibitem{Hsieh2008}
\bibinfo{author}{Hsieh, D.} \emph{et~al.}
\newblock \bibinfo{title}{A Topological Dirac insulator in a quantum spin hall
  phase}.
\newblock \emph{\bibinfo{journal}{Nature}} \textbf{\bibinfo{volume}{452}},
  \bibinfo{pages}{970--974} (\bibinfo{year}{2008}).
\newblock \urlprefix\url{http://dx.doi.org/10.1038/nature06843}.

\bibitem{Hsieh2009}
\bibinfo{author}{Hsieh, D.} \emph{et~al.}
\newblock \bibinfo{title}{Observation of unconventional quantum spin textures
  in topological insulators}.
\newblock \emph{\bibinfo{journal}{Science}} \textbf{\bibinfo{volume}{323}},
  \bibinfo{pages}{919--922} (\bibinfo{year}{2009}).
\newblock \urlprefix\url{http://www.sciencemag.org/content/323/5916/919.abstract}.

\bibitem{Fu2011}
\bibinfo{author}{Fu, L.}
\newblock \bibinfo{title}{Topological crystalline insulators}.
\newblock \emph{\bibinfo{journal}{Phys. Rev. Lett.}}
  \textbf{\bibinfo{volume}{106}}, \bibinfo{pages}{106802}
  (\bibinfo{year}{2011}).
\newblock \urlprefix\url{http://link.aps.org/doi/10.1103/PhysRevLett.106.106802}.

\bibitem{Kargarian2013}
\bibinfo{author}{Kargarian, M.} \& \bibinfo{author}{Fiete, G. A.}
\newblock \bibinfo{title}{Topological Crystalline Insulators in Transition Metal Oxides}.
\newblock \emph{\bibinfo{journal}{Phys. Rev. Lett.}}
 \textbf{\bibinfo{volume}{110}}, \bibinfo{pages}{156403}(\bibinfo{year}{2013}).
\newblock \urlprefix\url{http://link.aps.org/doi/10.1103/PhysRevLett.110.156403}.

\bibitem{Hsieh2014}
\bibinfo{author}{Hsieh, T. H.}, \bibinfo{author}{Liu, J.}, \& \bibinfo{author}{Fu, L.}
\newblock \bibinfo{title}{Topological crystalline insulators and Dirac octets in antiperovskites}.
\newblock \emph{\bibinfo{journal}{Phys. Rev. B}}
 \textbf{\bibinfo{volume}{90}}, \bibinfo{pages}{081112}(\bibinfo{year}{2014}).
\newblock \urlprefix\url{http://link.aps.org/doi/10.1103/PhysRevB.90.081112}.

\bibitem{Wan2011}
\bibinfo{author}{Wan, X.}, \bibinfo{author}{Turner, A.~M.},
  \bibinfo{author}{Vishwanath, A.} \& \bibinfo{author}{Savrasov, S.~Y.}
\newblock \bibinfo{title}{Topological semimetal and fermi-arc surface states in
  the electronic structure of pyrochlore iridates}.
\newblock \emph{\bibinfo{journal}{Phys. Rev. B}} \textbf{\bibinfo{volume}{83}},
  \bibinfo{pages}{205101--} (\bibinfo{year}{2011}).
\newblock \urlprefix\url{http://link.aps.org/doi/10.1103/PhysRevB.83.205101}.

\bibitem{Matsuura2013}
\bibinfo{author}{Matsuura, S.}, \bibinfo{author}{Chang, P.-Y.},
  \bibinfo{author}{Schnyder, A.~P.} \& \bibinfo{author}{Ryu, S.}
\newblock \bibinfo{title}{Protected boundary states in gapless topological
  phases}.
\newblock \emph{\bibinfo{journal}{New Journal of Physics}}
  \textbf{\bibinfo{volume}{15}}, \bibinfo{pages}{065001}
  (\bibinfo{year}{2013}).

\bibitem{Witczak-Krempa2014}
\bibinfo{author}{Witczak-Krempa, W.}, \bibinfo{author}{Chen, G.},
  \bibinfo{author}{Kim, Y.~B.} \& \bibinfo{author}{Balents, L.}
\newblock \bibinfo{title}{Correlated quantum phenomena in the strong spin-orbit
  regime}.
\newblock \emph{\bibinfo{journal}{Annu. Rev. Condens. Matter Phys.}}
  \textbf{\bibinfo{volume}{5}}, \bibinfo{pages}{57--82} (\bibinfo{year}{2014}).
\newblock \urlprefix\url{http://dx.doi.org/10.1146/annurev-conmatphys-020911-125138}.

\bibitem{Kargarian2011}
\bibinfo{author}{Kargarian, M.}, \bibinfo{author}{Wen, J.},
 \& \bibinfo{author}{Fiete, G. A.}
\newblock \bibinfo{title}{Competing exotic topological insulator phases in transition-metal oxides on the pyrochlore lattice with distortion}.
\newblock \emph{\bibinfo{journal}{Phys. Rev. B}}
 \textbf{\bibinfo{volume}{83}}, \bibinfo{pages}{165112}(\bibinfo{year}{2011}).
\newblock \urlprefix\url{http://link.aps.org/doi/10.1103/PhysRevB.83.165112}.


\bibitem{Yang2014}
\bibinfo{author}{Yang, B. -J.} \& \bibinfo{author}{Nagaosa, N.}
\newblock \bibinfo{title}{Emergent Topological Phenomena in Thin Films of Pyrochlore Iridates}.
\newblock \emph{\bibinfo{journal}{Phys. Rev. Lett.}}
 \textbf{\bibinfo{volume}{112}}, \bibinfo{pages}{246402}(\bibinfo{year}{2014}).
\newblock \urlprefix\url{http://link.aps.org/doi/10.1103/PhysRevLett.112.246402}.


\bibitem{Carter2012}
\bibinfo{author}{Carter, J.-M.}, \bibinfo{author}{Shankar, V.~V.},
  \bibinfo{author}{Zeb, M.~A.} \& \bibinfo{author}{Kee, H.-Y.}
\newblock \bibinfo{title}{Semimetal and topological insulator in perovskite
  iridates}.
\newblock \emph{\bibinfo{journal}{Phys. Rev. B}} \textbf{\bibinfo{volume}{85}},
  \bibinfo{pages}{115105} (\bibinfo{year}{2012}).
\newblock \urlprefix\url{http://link.aps.org/doi/10.1103/PhysRevB.85.115105}.

\bibitem{Zeb2012}
\bibinfo{author}{Zeb, M.~A.} \& \bibinfo{author}{Kee, H.-Y.}
\newblock \bibinfo{title}{Interplay between spin-orbit coupling and hubbard
  interaction in SrIro$_3$ and related $Pbnm$ perovskite oxides}.
\newblock \emph{\bibinfo{journal}{Phys. Rev. B}} \textbf{\bibinfo{volume}{86}},
  \bibinfo{pages}{085149} (\bibinfo{year}{2012}).
\newblock \urlprefix\url{http://link.aps.org/doi/10.1103/PhysRevB.86.085149}.

\bibitem{Cao2007}
\bibinfo{author}{Cao, G.} \emph{et~al.}
\newblock \bibinfo{title}{Non-Fermi-liquid behavior in nearly ferromagnetic SrIrO$_3$ single crystals}.
\newblock \emph{\bibinfo{journal}{Phys. Rev. B}} \textbf{\bibinfo{volume}{76}},
  \bibinfo{pages}{100402} (\bibinfo{year}{2007}).
\newblock \urlprefix\url{http://link.aps.org/doi/10.1103/PhysRevB.76.100402}.

\bibitem{Moon2008}
\bibinfo{author}{Moon, S. J.} \emph{et~al.}
\newblock \bibinfo{title}{Dimensionality-Controlled Insulator-Metal Transition and Correlated Metallic State in 5d Transition Metal Oxides Sr$_{n+1}$Ir$_n$O$3n+1$ (n=1, 2, and $\infty$ )}.
\newblock \emph{\bibinfo{journal}{Phys. Rev. Lett.}} \textbf{\bibinfo{volume}{101}},
  \bibinfo{pages}{226402} (\bibinfo{year}{2008}).
\newblock \urlprefix\url{http://link.aps.org/doi/10.1103/PhysRevLett.101.226402}.

\bibitem{Ran2010}
\bibinfo{author}{Ran, Y.}
\newblock \bibinfo{title}{Weak indices and dislocations in general topological
  band structures}.
\newblock \emph{\bibinfo{journal}{ArXiv e-prints 1006.5454}}
  (\bibinfo{year}{2010}).
\newblock \eprint{http://arxiv.org/abs/1006.5454}.

\bibitem{Ran2009}
\bibinfo{author}{Ran, Y.}, \bibinfo{author}{Zhang, Y.} \&
  \bibinfo{author}{Vishwanath, A.}
\newblock \bibinfo{title}{One-dimensional topologically protected modes in
  topological insulators with lattice dislocations}.
\newblock \emph{\bibinfo{journal}{Nat. Phys.}} \textbf{\bibinfo{volume}{5}},
  \bibinfo{pages}{298--303} (\bibinfo{year}{2009}).
\newblock \urlprefix\url{http://dx.doi.org/10.1038/nphys1220}.

\bibitem{Kitaev2009}
\bibinfo{author}{Kitaev, A.}
\newblock \bibinfo{title}{Periodic table for topological insulators and
  superconductors}.
\newblock \emph{\bibinfo{journal}{AIP Conf. Proc.}}
  \textbf{\bibinfo{volume}{1134}}, \bibinfo{pages}{22--30}
  (\bibinfo{year}{2009}).
\newblock \urlprefix\url{http://dx.doi.org/10.1063/1.3149495}.

\bibitem{Schnyder2008}
\bibinfo{author}{Schnyder, A.~P.}, \bibinfo{author}{Ryu, S.},
  \bibinfo{author}{Furusaki, A.} \& \bibinfo{author}{Ludwig, A. W.~W.}
\newblock \bibinfo{title}{Classification of topological insulators and
  superconductors in three spatial dimensions}.
\newblock \emph{\bibinfo{journal}{Phys. Rev. B}} \textbf{\bibinfo{volume}{78}},
  \bibinfo{pages}{195125} (\bibinfo{year}{2008}).
\newblock \urlprefix\url{http://link.aps.org/doi/10.1103/PhysRevB.78.195125}.

\bibitem{Yang2011}
\bibinfo{author}{Yang, K.-Y.}, \bibinfo{author}{Lu, Y.-M.} \&
  \bibinfo{author}{Ran, Y.}
\newblock \bibinfo{title}{Quantum hall effects in a weyl semimetal: Possible
  application in pyrochlore iridates}.
\newblock \emph{\bibinfo{journal}{Phys. Rev. B}} \textbf{\bibinfo{volume}{84}},
  \bibinfo{pages}{075129} (\bibinfo{year}{2011}).
\newblock \urlprefix\url{http://link.aps.org/doi/10.1103/PhysRevB.84.075129}.

\bibitem{Longo1971}
\bibinfo{author}{Longo, J.}, \bibinfo{author}{Kafalas, J.} \&
  \bibinfo{author}{Arnott, R.}
\newblock \bibinfo{title}{Structure and properties of the high and low pressure
  forms of SrIrO$_3$}.
\newblock \emph{\bibinfo{journal}{Journal of Solid State Chemistry}}
  \textbf{\bibinfo{volume}{3}}, \bibinfo{pages}{174--179}
  (\bibinfo{year}{1971}).
\newblock \urlprefix\url{http://dx.doi.org/10.1016/0022-4596(71)90022-3}.

\bibitem{Zhao2008}
\bibinfo{author}{Zhao, J.~G.} \emph{et~al.}
\newblock \bibinfo{title}{High-pressure synthesis of orthorhombic SrIrO$_3$
  perovskite and its positive magnetoresistance}.
\newblock \emph{\bibinfo{journal}{Journal of Applied Physics}}
  \textbf{\bibinfo{volume}{103}}, \bibinfo{pages}{103706} (\bibinfo{year}{2008}).
\newblock \urlprefix\url{http://ieeexplore.ieee.org/stamp/stamp.jsp?tp=&arnumber=4945811&isnumber=4945743}.


\bibitem{Heeger1988}
\bibinfo{author}{Heeger, A. J.} \emph{et~al.}
\newblock \bibinfo{title}{Solitons in conducting polymers}.
\newblock \emph{\bibinfo{journal}{Rev. Mod. Phys.}}
  \textbf{\bibinfo{volume}{60}}, \bibinfo{pages}{781--850} (\bibinfo{year}{1988}).
\newblock \urlprefix\url{http://link.aps.org/doi/10.1103/RevModPhys.60.781}.


\bibitem{Matsuno2014}
\bibinfo{author}{{Matsuno}, J.} \emph{et~al.}
\newblock \bibinfo{title}{{Engineering spin-orbital magnetic insulator by
  tailoring superlattices}}.
\newblock \emph{\bibinfo{journal}{ArXiv e-prints}}  (\bibinfo{year}{2014}).
\newblock \eprint{http://arxiv.org/abs/1401.1066}.

\bibitem{Takagi14}D. Hirai and H. Takagi, private communication.


\bibitem{Wen2012}
\bibinfo{author}{Wen, X.-G.}
\newblock \bibinfo{title}{Symmetry-protected topological phases in
  noninteracting fermion systems}.
\newblock \emph{\bibinfo{journal}{Phys. Rev. B}} \textbf{\bibinfo{volume}{85}},
  \bibinfo{pages}{085103} (\bibinfo{year}{2012}).
\newblock \urlprefix\url{http://link.aps.org/doi/10.1103/PhysRevB.85.085103}.


\bibitem{Morimoto2013}
\bibinfo{author}{Morimoto, T.} \& \bibinfo{author}{Furusaki, A.}
\newblock \bibinfo{title}{Topological classification with additional symmetries
  from Clifford algebras}.
\newblock \emph{\bibinfo{journal}{Phys. Rev. B}} \textbf{\bibinfo{volume}{88}},
  \bibinfo{pages}{125129} (\bibinfo{year}{2013}).
\newblock \urlprefix\url{http://link.aps.org/doi/10.1103/PhysRevB.88.125129}.

\bibitem{Lu2014a}
\bibinfo{author}{{Lu}, Y.-M.} \& \bibinfo{author}{{Lee}, D.-H.}
\newblock \bibinfo{title}{{Inversion symmetry protected topological insulators
  and superconductors}}.
\newblock \emph{\bibinfo{journal}{ArXiv e-prints}}  (\bibinfo{year}{2014}).
\newblock \eprint{http://arxiv.org/abs/1403.5558}.

\bibitem{Yi2014}
\bibinfo{author}{Yi, H.} \emph{et~al.}
\newblock \bibinfo{title}{Evidence of Topological Surface State in Three-Dimensional Dirac Semimetal Cd$_3$As$_2$}.
\newblock \emph{\bibinfo{journal}{Sci. Rep.}} \textbf{\bibinfo{volume}{4}},
 \bibinfo{pages}{6106} (\bibinfo{year}{2014}).
\newblock \urlprefix\url{http://dx.doi.org/10.1038/srep06106}.

\bibitem{Schnyder2011}
\bibinfo{author}{Schnyder, A.~P.} \& \bibinfo{author}{Ryu, S.}
\newblock \bibinfo{title}{Topological phases and surface flat bands in
  superconductors without inversion symmetry}.
\newblock \emph{\bibinfo{journal}{Phys. Rev. B}} \textbf{\bibinfo{volume}{84}},
  \bibinfo{pages}{060504} (\bibinfo{year}{2011}).
\newblock \urlprefix\url{http://link.aps.org/doi/10.1103/PhysRevB.84.060504}.

\end{thebibliography}

\newpage

\begin{table*}
\centering
\begin{tabular}{|c|c||c|c|c|}
\hline
mirror $\Pi_m$&chiral $\mathcal{C}$&Classification&Topological invariants&Surface zero modes\\
\hline
Yes&Yes&$\mbz\times\mbz$&$(W^+,W^-)=(1,-1)$&Yes ($k_c=\frac\pi c$)\\
\hline
No&Yes&$\mbz$&$W\equiv W^++W^-$& Yes ($k_c=\frac\pi c+\delta,~~|\delta|\ll1$)\\
\hline
Yes&No&0&-&No\\
\hline
No&No&0&-&No\\
\hline
 \end{tabular}
\caption{{\bf Classification of topological crystalline metals and possible surface states on a generic side surface parallel to $\hat{\bf z}$-axis.} Note that $[\Pi_m,\mathcal{C}]=0$.
$(W^+,W^-)$ is the pair of winding numbers obtained in $\Pi_m=\pm1$ subspaces of block-diagonalized Hamiltonian.}
\label{tab:classification:[xy0]}
\end{table*}

\begin{table*}[h]
\centering
\begin{tabular}{|c||c|c|c|c||c|c|}
\hline
B field & b-glide $\Pi_b$ & n-glide $\Pi_n$ & mirror $\Pi_m$ & chiral $\mathcal{C}$ & Fermi surfaces & Zero-energy surface states\\
\hline
$[1\bar{1}0]$ & No & Yes & No & Yes & Nodal rings & Flat bands at $k_c=\frac{\pi}c\pm\delta$ \\
\hline
$[110]$ & Yes & No & No & Yes & Nodal points & Fermi arcs between Weyl points \\
\hline
$[001]$ & No & No & Yes & No & Gapped & No \\
\hline
\end{tabular}
\caption{{\bf Various TR-breaking perturbations and consequences on the nodal ring and surface states (e.g. on $[1\bar{1}0]$ side surface).} When the magnetic field is along $[110]$ direction, the nodal ring splits into a pair of Dirac nodes along ${\rm U} \rightarrow {\rm R}$. On the other hand, the nodal ring is completely gapped if the magnetic field has a nonzero $\hat{\bf z}(\hat{\bf c})$-direction component.}
\label{tab:dp}
\end{table*}

\begin{figure}
\centering
\includegraphics[scale=0.6]{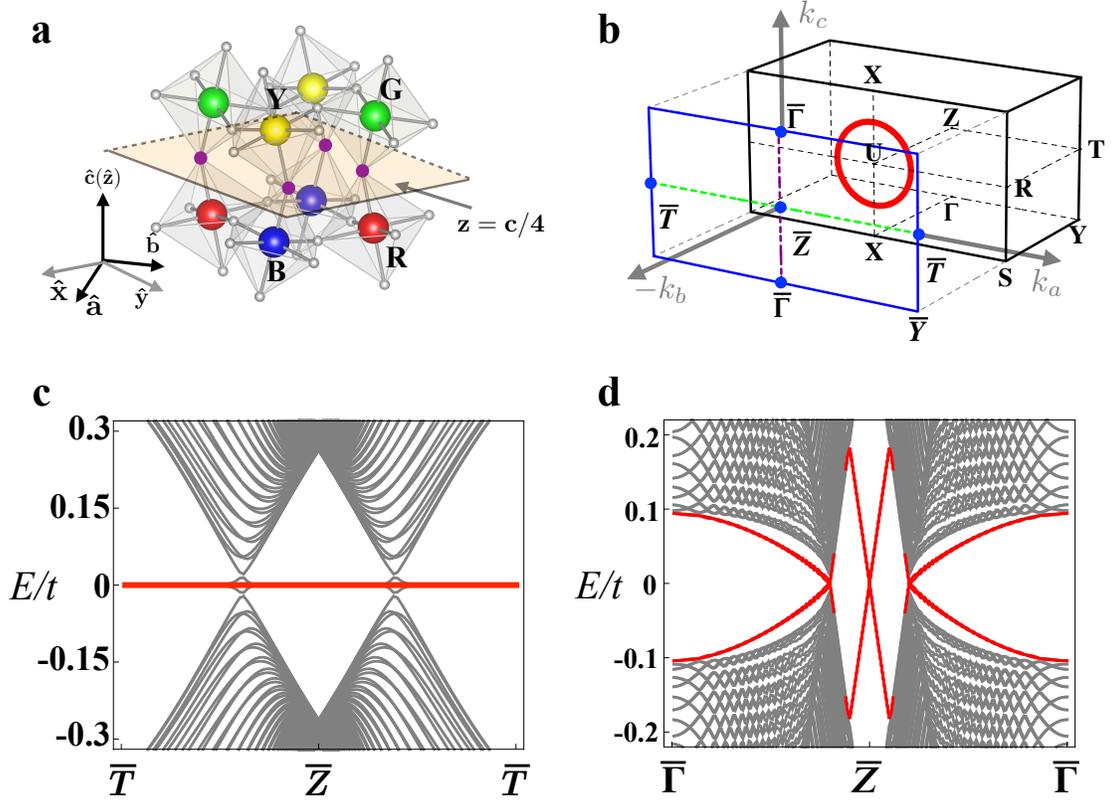}
\caption{{\bf Crystal structure and surface states on $[1\bar{1}0]$ side plane.} (a)Crystal structure of AIrO$_3$ with $\hat{\bf a}$-axis along $(110)$, $\hat{\bf b}$-axis along $(1\bar{1}0)$ and $\hat{\bf c}$-axis along $\hat{\bf z}$-direction, respectively. The unit cell contains four Ir atoms: blue (B), red (R), yellow (Y) and green (G) represent different oxygen octahedra environment. The length of Bravais lattice unit vectors $\hat{\bf a}/\hat{\bf b}$ and $\hat{\bf c}$ axis are $a/a$ and $c$, respectively. A mirror symmetry plane (colored with light orange), mapping $z$ to $-z$, locates at $z=\frac{c}4$ where four oxygen atoms (purple solid circles) are within the same plane. (b) Special $k$-points in the 3D bulk BZ and 2D surface BZ for $[1\bar{1}0]$ surface. The location of the nodal ring depict as red circles on U-R-S-X plane in 3D BZ.
Slab-geometry surface spectrum (c) for $[1\bar{1}0]$ surface at $k_c=\frac\pi c$, plotted along the high symmetry line (green dashed line) $\bar{T}\rightarrow\bar{Z}\rightarrow\bar{T}$, and (d) for $[1\bar{1}0]$ surface at $k_a=0$ by varying $k_c$ to follow $\bar{\Gamma}\rightarrow\bar{Z}\rightarrow\bar{\Gamma}$ (purple dashed line), where red lines represent surface states.}
\label{fig:pd}
\end{figure}

\begin{figure}
\centering
\includegraphics[scale=0.6]{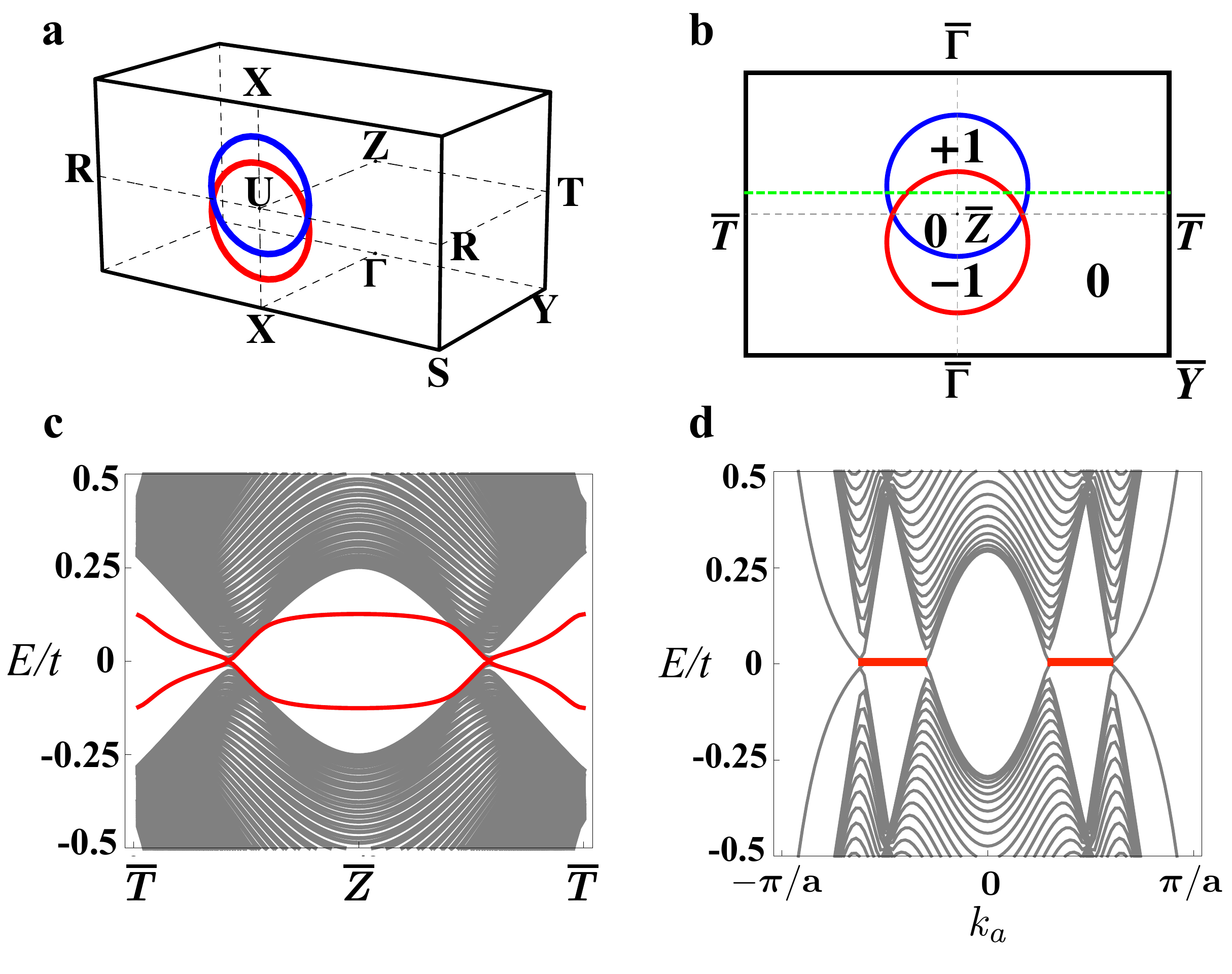}
\caption{{\bf Evolution of the nodal ring under various TR-breaking terms and corresponding surface modes.} (a) When the magnetic field ${\hat {\bf h}} \parallel {\hat{\bf b}}$, a pair of nodal rings contain one ring (blue) shifted upwards along $\hat{\bf c}(\hat{\bf z})$-axis and the other one (red) shifted  downwards.  (b) The corresponding winding number $W_{[1\bar{1}0]}$ distribution on $\bar{Z}$-$\bar{T}$-$\bar{Y}$-$\bar{\Gamma}$ plane.  Slab spectrum (c) when  $k_c=\frac\pi c$ for $[1\bar{1}0]$ surface plotted as a function of $k_a$ along the high symmetry line $\bar{T}\rightarrow\bar{Z}\rightarrow\bar{T}$, and  (d) when $k_c=\frac\pi c + \delta$ indicated by the green dashed line in (b). }
\label{fig:trb}
\end{figure}

\begin{figure}
\centering
\includegraphics[scale=0.6]{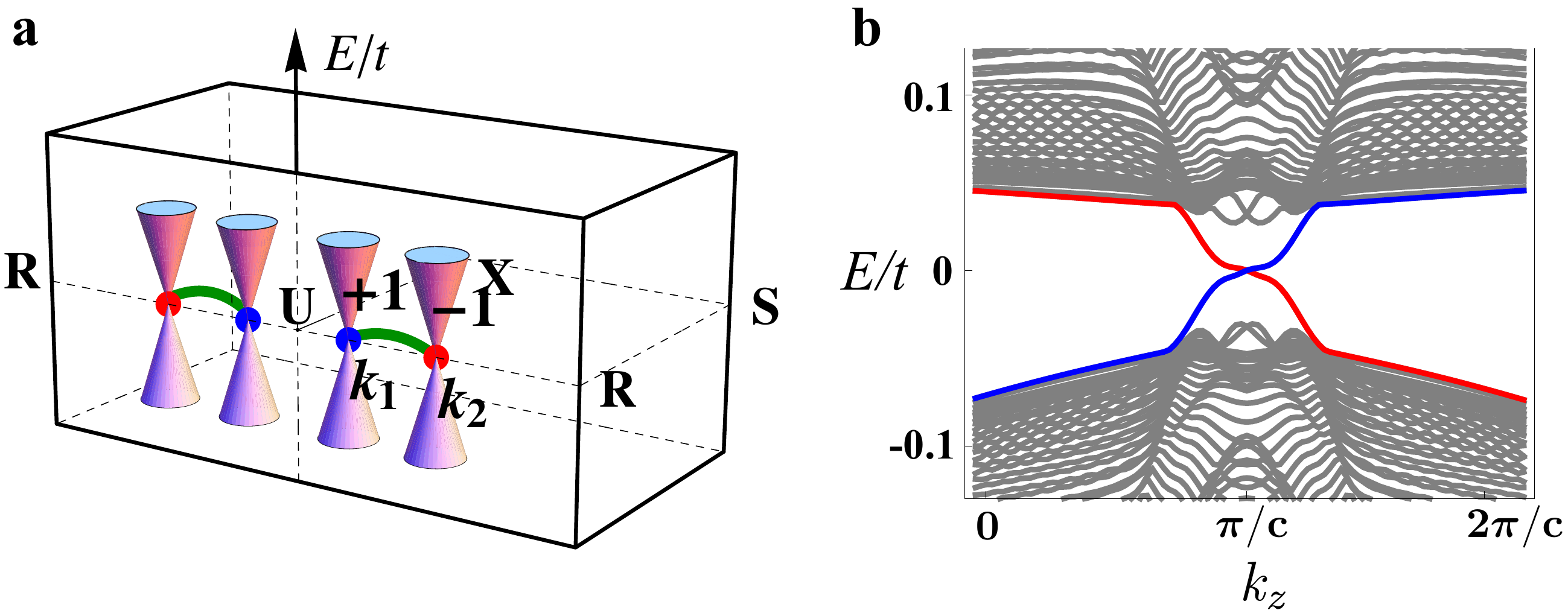}
\caption{{\bf Emergence of Weyl fermions on $[1\bar{1}0]$ side surface.}  (a) Two pairs of Weyl node located along the high symmetry line ${\rm U} \rightarrow {\rm R}$ at ${\bf k}_1$ and ${\bf k}_2$, respectively. One of the Weyl fermion (blue) has $+1$ chirality but the other Weyl node (red) has opposite chirality. The Fermi arc connecting those two Weyl nodes is colored by green. (b)The edge states spectra when $k_a=\frac{0.7}{a}$ which is in between two Weyl nodes plotted as a function of $k_c$.}
\label{fig:w1}
\end{figure}

\begin{figure}
\centering
\includegraphics[scale=0.6]{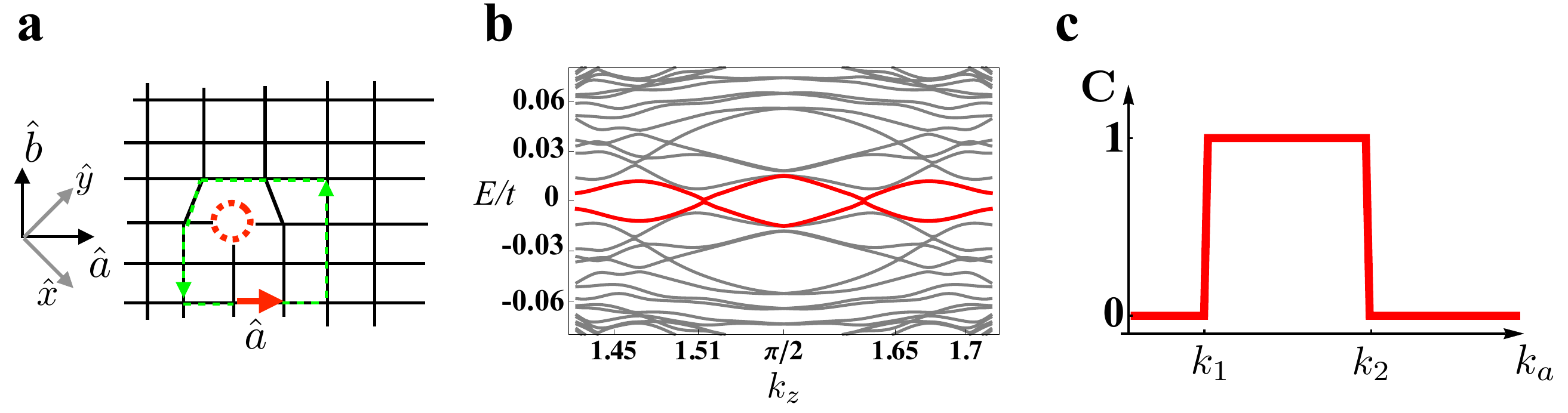}
\caption{{\bf Disloction spectrum and Chern number.} (a) Top view of an edge dislocation with Burgers vector $\hat{\bf a}$ (colored by red arrow), whose dislocation line is along $\hat{\bf c}$ axis. Dashed green line denotes a trajectory around the dislocation core (red circle). Clearly when we go around the dislocation once, we need an extra translation by the Burgers vector (the red arrow), equal to Bravais primitive vector $\hat{\bf a}$ in this case) to return to the starting point. (b) The dislocation spectrum for system size $39 \times 15$ contains two pairs of gapless helical modes (highlighted with red color) localized at each dislocation line. (c) Chern number as a function of $k_a$ for all occupied bands. Opposite Chern number jump  indicates the Weyl fermion at ${\bf k}_1$ and ${\bf k}_2$ has different chirality.}
\label{fig:dis}
\end{figure}


%

\end{document}